\documentclass[journal=ancac3,manuscript=article,
layout=twocolumn
]{achemso}

\usepackage[T1]{fontenc} 
\usepackage{xcolor}

\newcommand{\NEELAddress}{Univ. Grenoble-Alpes, CNRS, Grenoble INP, Inst. NEEL, Grenoble, France}
\newcommand{\IRIGAddress}{Univ. Grenoble Alpes, CEA, Grenoble INP, IRIG, PHELIQS, Grenoble, France }
\newcommand{\CNNAddress}{Univ. Paris-Saclay, CNRS, Centre for Nanoscience and Nanotechnology, Palaiseau, France}
\author{Edith Bellet-Amalric}
\affiliation{\IRIGAddress}
\author{Federico Panciera}
\author{Gilles Patriarche}
\author{Laurent Travers}
\affiliation{\CNNAddress}
\author{Martien den Hertog}
\affiliation{\NEELAddress}
\author{Jean-Christophe Harmand}
\author{Frank Glas}
\affiliation{\CNNAddress}
\author{Jo\"{e}l Cibert}
\email{joel.cibert@neel.cnrs.fr}
\affiliation{\NEELAddress}

\title[in-situ VSS]
  {Regulated dynamics with two-monolayer steps in vapor-solid-solid growth of nanowires}

\abbreviations{}

\keywords{\emph{nanowires, layer-by-layer growth, epitaxial growth, II-VI semiconductors, in-situ growth, transmission electron microscopy, vapor-solid-solid}}

\begin{document}








\begin{abstract}
\textbf{The growth of ZnTe nanowires and ZnTe-CdTe nanowire heterostructures is studied by \emph{in situ} transmission electron microscopy. We describe the shape, and the change of shape, of the solid gold nanoparticle during vapor-solid-solid growth. We show the balance between one-monolayer and two-monolayer steps which characterizes the vapor-liquid-solid and vapor-solid-solid growth modes of ZnTe. We discuss the role of the mismatch strain and lattice coincidence between gold and ZnTe on the predominance of two-monolayer steps during vapor-solid-solid growth, and on the subsequent self-regulation of the step dynamics. Finally, the formation of an interface between CdTe and ZnTe is described.}
\end{abstract}


\vspace{\baselineskip}
\vspace{\baselineskip}

Semiconductor nanowires are at the core of an intense interdisciplinary research (see Ref.~\citenum{Garnett2019} and references therein) involving chemistry, materials science, physics, and electrical engineering, with applications in nanoelectronics \cite{Jia2019}, laser and photonic circuits \cite{Gu2021}, biology and medicine, energy \cite{Goktas2018} including photovoltaics \cite{Consonni2019}, and quantum technologies from quantum transport in hybrid systems \cite{Prada2020} to single photon emitters \cite{Arakawa2020}. Nanowires also constitute a model system to improve our understanding of the growth of crystalline nanostructures \cite{Harmand2018}.

It is widely admitted that the growth of a nanowire takes place through layer-by-layer growth at the interface between the seed and the nanowire \cite{Dubrovskii2004,Glas2013}. \emph{In-situ} TEM studies - images and movies recorded in a Transmission Electron Microscope equipped with a source of material allowing the growth of the nanowire- have confirmed this idea, and brought details of step dynamics for various materials. The most studied are silicon \cite{Ross2010,Hofmann2008,Wen2010PRL,Wen2010NL,Chou2016,Ngo2021}, and III-Vs \cite{Harmand2018,Jacobsson2016,Maliakkal2019Nat,Panciera2020,Maliakkal2020,Maliakkal2021arX}, including nitrides \cite{Gamalski2016}. The movies identify two periods which constitute the growth cycle: the nucleation of 1~ML steps at the liquid-solid interface, and the expansion of the island thanks to the propagation of the step across the interface. As the nucleation of the steps is a random process, it raises the question of antibunching \cite{Glas2010}, particularly if the velocity of the steps is large, so that the propagation time is much shorter than the nucleation time. More recent studies have focused onto the role of the post-nucleation period where the step propagation may be limited by refilling \cite{Dubrovskii2017,Glas2020}.

Most of these studies deal with the so-called Vapor-Liquid-Solid (VLS) growth mode, \emph{i.e.}, the seed is a liquid nanodroplet. An alternative is the Vapor-Solid-Solid (VSS) growth mode, where the seed is a solid nanoparticle \cite{Persson2004,Dick2005a,Dick2005b,YewuWang2006,Krogstrup2009,Wen2009,Dick2012,Kodambaka2016}. VSS is thought to be particularly attractive in order to achieve sharp interfaces when a quantum dot is inserted in a nanowire, thanks either to a smaller growth rate, or to a potentially smaller reservoir effect (the content of the constituents of the nanowire heterostructure being smaller in the nanoparticle than in the nanodroplet) . Specific models have been developed for VSS \cite{Golovin2008,Cui2015}, the most recent ones taking into account the lattice mismatch and possible lattice coherence at the interface \cite{Koryakin2019a,Koryakin2019b}. From the experimental point of view, VSS was explored by in-situ TEM for silicon (with Cu \cite{Wen2010NL} or AuAg nanoparticles \cite{Chou2016}, or double-phase Cu-Sn seed \cite{Ngo2021}) and for GaAs with a gold seed \cite{Maliakkal2021arX}. In the latter case, a mixing of 1~ML and 2~ML steps was observed in VSS while only 1~ML were present in VLS.

Distinguishing VLS and VSS thanks to post-growth studies performed at room temperature is not an easy task. The observation, or not, of facets can be misleading. For instance, the role of faceting in the growth of InAs NWs was discussed in the context of the VLS growth mode \cite{Lin2012}. Conversely, crystalline nanoparticles can feature a quasi-spherical shape, as shown below. Even, VSS was argued to be vapour-quasisolid-solid \cite{Mohammad2009}. And of course, the effect of the growth mode on the formation of an axial heterostructure such as a quantum dot needs to be evaluated by \emph{in-situ} TEM. Such studies are scarce: The Si-Ge interface was studied quite early \cite{Wen2009} with a resolution too low to evidence the interface steps. For III-Vs, this would require using more than two effusion cells. A good opportunity is offered by the congruent nature of the sublimation of some II-VI semiconductors:  Fabricating a ZnTe-CdTe interface requires only two effusion cells.

The growth of II-VI nanowires is less developed than that of III-V nanowires. The II-VI semiconductors offer specific properties, such as an efficient coupling to light for photovoltaic and lighting applications. The strong electron-hole interaction gives rise to a large exciton-biexciton splitting, so that single photon emission was demonstrated up to room temperature in CdSe quantum dots embedded in a ZnSe nanowire \cite{Bounouar2012}. This dot-nanowire configuration in CdTe-ZnTe allows one to adjust the built-in strain in order to manipulate the magnetic properties introduced through Mn impurities \cite{Wojnar2012,Szymura2015} or to control the light-hole heavy-hole character of the ground state \cite{Jeannin2017} for an increased flexibility of the optical and spin properties.

ZnTe nanowires can be grown by MBE using a gold seed, on GaAs or Si substrates \cite{Janik2006,Janik2007,Wojnar2013}, where the growth mode is assumed to be VLS due to the formation of a eutectic of gold with gallium or silicon, or on ZnTe buffer layer \cite{Rueda2014}, where the growth mode is assumed to be VSS. The insertion of a II-VI buffer layer was exploited recently in order to guarantee a VSS growth of ZnS nanowires \cite{Kumar2021}. In the VSS case, ex-situ TEM reveals facetted nanoparticles, but also quasi-spherical shapes \cite{Rueda2016} which may suggest VLS, or even bulb-like shapes \cite{Orru2018}. Simple models based on the diffusion along the nanowire sidewalls describe satisfactorily the measured growth rates, with a diffusion length of the order of 100 nm \cite{Rueda2016}, provided one takes into account a change in the shape of the nanoparticle (and of the nanoparticle-nanowire contact area) during the growth \cite{Rueda2016}. There is no evidence of a role of the nucleation of islands at the Au-ZnTe interface. Finally, the interfaces of a CdTe dot inserted in a ZnTe nanowire grown on a ZnTe buffer layer evidence a reservoir effect compatible with an average Zn/Cd content of the nanoparticle around 2\% \cite{Orru2018}, much smaller than when the nanowires are grown on GaAs \cite{Kirmse2009}. Our present knowledge of the growth of II-VI nanowires and nanowire heterostructures relies upon post-growth, \emph{ex-situ} studies, and phenomenological interpretations: A better understanding calls for \emph{in-situ} observations.

\begin{figure*}
 \includegraphics[width=2\columnwidth]{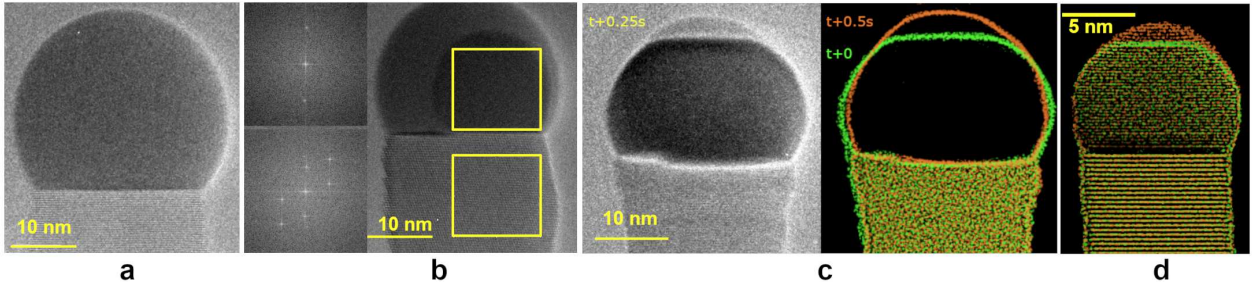}
  \caption{ Different morphologies of the gold seed on the ZnTe nanowire. (a) liquid gold droplet (NW 13-25, Si substrate). (b) a solid core in a liquid droplet, with FFTs of the squared areas (NW Frame6-25, Si substrate). (c) solid nanoparticle, with a step coming from the left (sequence 44-26, Si substrate); The left image is the intermediate image at time $t+0.25~$s; The right image shows the superposition of two images, colored in green at time $t$ and red at $t+0.5~$s. (d) solid nanoparticle, two superimposed frames, colored in green at time $t$ and red at $t+0.25~$s, with a step coming from the back (NW 31-06, SiC substrate.)} \label{fig1}
\end{figure*}

In the present study, we observe VLS and VSS growth of ZnTe and ZnTe/CdTe nanowires in real time in a modified environmental TEM \cite{Harmand2018}, \cite{Panciera2020}. Growth by molecular beam epitaxy with the stoichiometric flux from a ZnTe and a CdTe cell is catalyzed by Au particles dispersed on a heating carrier substrate which consists of a thin patterned substrate made either of silicon or silicon carbide.

\section{Results and discussion}

Results are presented on the shape and nature of the gold seed, the layer-by-layer growth of the nanowire, and the formation of a CdTe-ZnTe interface.

\subsection{Solid nanoparticle \emph{vs.} liquid nanodroplet}

This section focuses onto the phase of the gold seed - solid nanoparticle or liquid droplet - and its shape.

The face-centered cubic (fcc) structure of the gold nanoparticle and its orientation with respect to ZnTe nanowires grown under standard conditions \cite{Rueda2014} were confirmed \emph{ex situ} by X-ray diffraction from an ensemble of as-grown nanowires (Supporting Information Sup1a). Several orientations are observed upon dewetting a sub-monolayer gold film on a ZnTe layer, all with the $<111>$ axes parallel in Au and ZnTe, but with different orientations within the $(111)$ plane. Similar orientations have been found for Au nanoparticles on Si \cite{Daudin2012}. However, as soon as the growth of the ZnTe nanowires has started, only nanoparticles oriented parallel to the nanowire, with a common orientation within the $(111)$ plane, are observed. With the lattice parameter of ZnTe, $a_0^{ZnTe}=0.610~$nm, and that of Au, $a_0^{Au}=0.408~$nm, there is a good coincidence at the interface with $2 a_0^{ZnTe}⁄(3 a_0^{Au} )=(1+f)$, $f=-0.3\%$. The coincidence is achieved thanks to a small compression of the Au lattice at the interface. The small coincidence mismatch and the induced strain, as well as the lattice parameter change which could be induced by the incorporation of a few percent of Cd or Zn into the gold seed \cite{Owen1945}, are too small to be detected from TEM on such small nanoparticles (see Supplementary information Sup1a). Note that we expect no Te accumulation, the maximum content being estimated to be as low as $0.15\%$ \cite{Okamoto1984}\cite{JianhuaWang2006}.

We turn now to the \emph{in situ} TEM during the gold-seeded growth of ZnTe nanowires on two types of substrates, silicon and silicon carbide.

As expected, we do not observe any amorphous layer at the surfaces of the nanowire and the seed. This layer, which is present in \emph{ex situ} TEM images of Au-seeded ZnTe \cite{Rueda2014} or ZnO nanowires \cite{Simon2013}, is thus confirmed to be of post-growth origin.

\begin{figure*}
 \includegraphics[width=2\columnwidth]{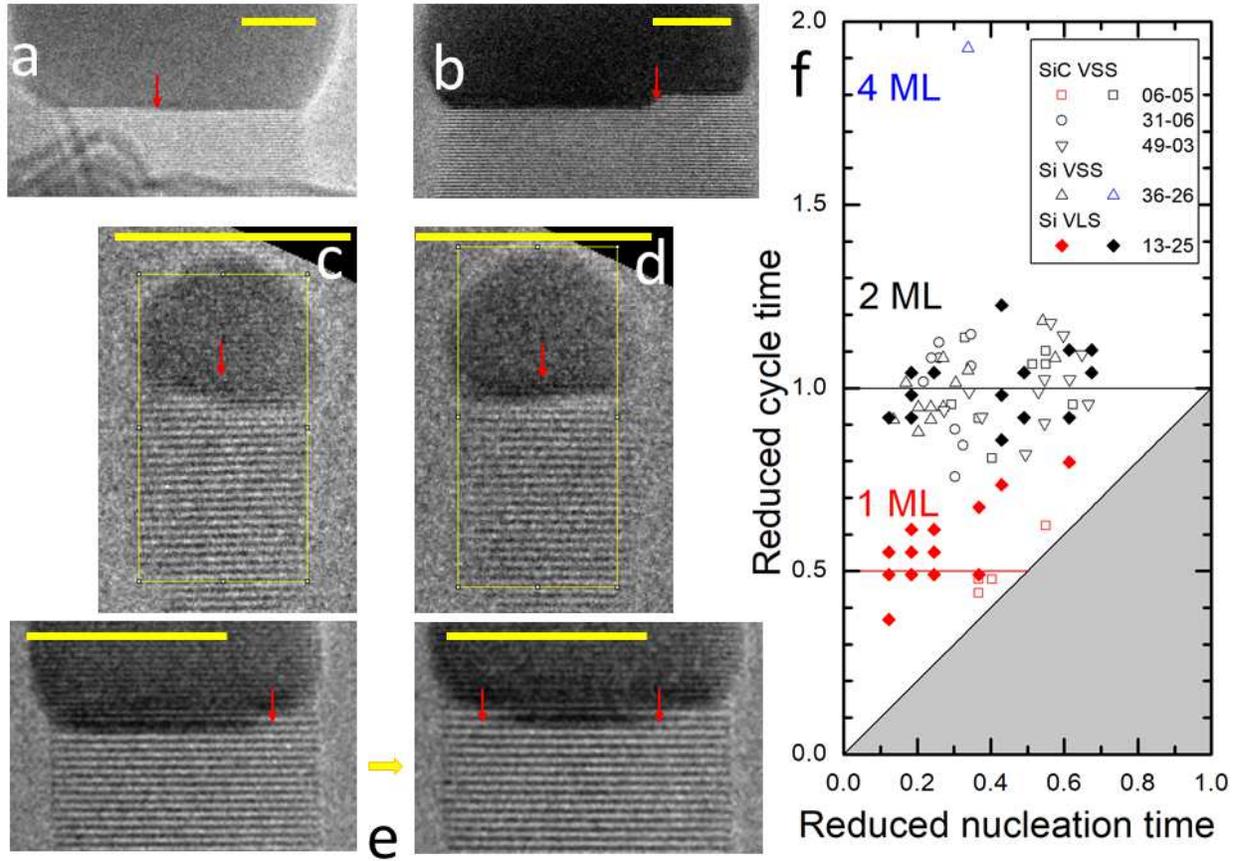}
  \caption{ (a) a 1~ML step in VLS (NW 13-25, Si substrate); (b) a 4~ML step in VSS (NW 35-26, Si substrate). (c) a 2~ML step in VSS (NW 06-05, SiC substrate); the yellow rectangle shows the area for the calculation of the FFT in Fig.~\ref{fig3}. (d) a 1~ML step in VSS (same nanowire). (e) a 2~ML step in VSS, which started from the side (left image) and finally (right image) reached the back- or front-side after changing direction (NW 31-06, SiC substrate). All steps but the last one propagate regularly from one side of the nanowire to the opposite one. All scale bars are 10 nm long. (f) Cycle time as a function of the nucleation time, for different nanowires as indicated. All times for a given nanowire are normalized to the cycle time of 2~ML steps averaged for this nanowire. Different symbols identify the data for a given nanowire; solid symbols are for VLS, open symbols for VSS, red for 1~ML steps, black fot 2~ML and blue for 4~ML.} \label{fig2}
\end{figure*}

On the silicon substrate, we observed liquid spherical droplets with a diameter in the 20 to 40 nm range after dewetting and growth start (see Fig.~\ref{fig1}a and Movie 1 in Supporting Information). This size is of the same order of magnitude as in standard MBE on a Si or a GaAs buffer layer \cite{Janik2007}). Gold is known to form a eutectic with $18.6\%$ Si at $363^\circ$C. Upon cooling down we obtained solid nanoparticles (Fig.~\ref{fig1}c and Movie 2 in Supporting Information). We sometimes observed the formation of two separate crystals (not shown). The values of the lattice parameters of these two crystals (and the observation of a corresponding moir\'{e} pattern  when they are superimposed) are compatible with almost pure Au and Si. This is compatible with the maximum solubility of Si in solid Au, around $2\%$ \cite{Okamoto-Massalski1983}. These nanoparticles feature complex facetted shapes and both may form a part of the interface with the ZnTe nanowire. An interesting intermediate case is the formation of a core which appears darker than the surrounding liquid AuSi droplet; It is crystalline, as shown in Fig.~\ref{fig1}b with the Fast Fourier Transforms (FFT) in the inset. An intriguing feature is the fact that the solid core moves rapidly inside the liquid droplet, in coincidence with the steps at the interface, as exemplified in Movie 3 of Supporting Information. The presence of such a solid-liquid interface rapidly moving in a dual-phase Cu-Sn seed, and its role on the growth of the nanowire, were also demonstrated by \emph{in-situ} TEM \cite{Ngo2021} during the growth of Si nanowires. A complete series, liquid droplet / solid core / bicrystal, was reported for Au-Ge \cite{Kim2012} and the complex pathways it induces for the growth of the nanowire were discussed. And solid nickel silicides were introduced on purpose into the droplet of VLS growth of silicon nanowires and could be used to fabricate heterostructures \cite{Panciera2015}.

With the silicon carbide substrate (Fig.~\ref{fig1}d), solid gold nanoparticles were systematically observed, with a diameter in the 6 to 20 nm range, \emph{i.e.}, the same order of magnitude as in standard MBE on a ZnTe buffer layer \cite{Rueda2016}. The Au $(111)$ on ZnTe $(111)$ orientation was confirmed on most nanowires. Precisely orienting the nanowire along a given zone axis is time-consuming and the nanowire sidewalls deteriorate under the electron beam. Therefore, we preferred to focus onto the growth dynamics and typically chose a NW for observation if the $(111)$ plane was visible, without orienting the nanowire on a crystallographic zone axis.

The gold particle can exhibit clear lattice planes over the whole volume, and at the same time assume different shapes, from quasi spherical to strongly facetted.

Rapid shape changes were even observed during the nanowire growth. Figures~\ref{fig1}c and \ref{fig1}d show two examples of such a dramatic change from facets to sphere during purely VSS growth, on Si or on SiC substrates  (Fig.~\ref{fig1}c and \ref{fig1}d respectively), see also Movie 4 in Supporting Information). The change takes place between two frames, in less than 0.2~s. In both cases, a step crosses the structure at the nanowire/nanoparticle interface. Note that the crystal structure is observed over the whole gold nanoparticle in Fig.~\ref{fig1}c-d. A significant amount of material is displaced during the change of shape, from the periphery to the apex. However, the corresponding change of the total volume of the nanoparticle is too small to be measured, and at most of the order of 1~ML over the surface. The change of shape takes place at constant volume.

A first, simple conclusion of this section is that the shape of the gold seed is not a good indication of its phase, liquid or solid. In the following, we will thus confirm the VLS or VSS nature of the growth from the observation of gold lattice planes.

\subsection{Layer by layer growth}

This section describes the behavior of the steps which form at the gold-semiconductor interface and ensure the growth of the nanowire. We define the monolayer (ML) in ZnTe as a set of two atomic planes, Zn and Te. Our three main observations are (1) the preponderance of 2~ML steps in VSS, which we tentatively ascribe to the 2:3 lattice coincidence at the interface, (2) a remarkable, although not complete, self-regulation of the step propagation, and (3) a deviation from the simple linear increase of the island area with time, even for the steps exhibiting a smooth motion from one side to the opposite side across the interface.

During the growth of the ZnTe nanowires, we observed 1~ML ZnTe steps, 2~ML steps, and even (few) 4~ML steps. Figure~\ref{fig2} shows examples of the different cases with solid or liquid gold. From 95 well-characterized steps present in Fig.~\ref{fig2}f:
\begin{itemize}
  \item In VSS, either on SiC or on Si substrate, more than 90\% are 2~ML steps, very few 1~ML,
  \item In VLS, on Si substrate, 40\% are 1~ML and 60\% are 2~ML steps.
\end{itemize}

\begin{figure}
 \includegraphics[width=\columnwidth]{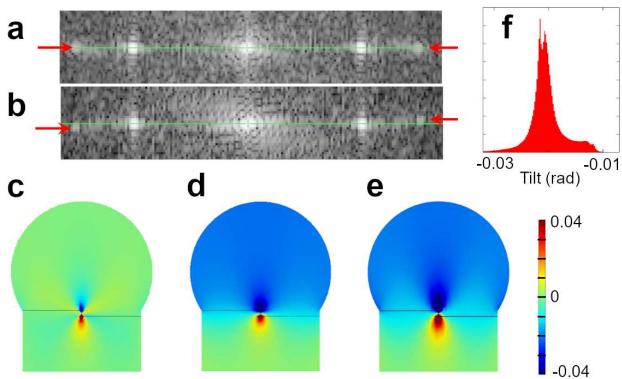}
  \caption{ (a) FFT of the nanowire + nanoparticle with a 2~ML step (image c of Fig.~\ref{fig2}, area marked by the yellow rectangle); The coincidence of the green line and the red arrows shows that the Au and ZnTe (111) lattice planes are parallel. (b) FFT of the nanowire + nanoparticle with a 1~ML step (image d of Fig.~\ref{fig2}, area marked by the yellow rectangle); green line and red arrows are shifted, which demonstrates the tilt of the gold lattice with respect to the ZnTe one (same NW); (c) to (e) for a 1ML step, map of the calculated shear strain (c), rotation (d) and tilt of (111) planes (e); (f) histogram of the tilt. The dimensions are from NW 06-05: The radius of the nanowire is 3.6 nm, the radius of the nanoparticle 4.3 nm and its height 6.4 nm.} \label{fig3}
\end{figure}

An effect of the lattice mismatch along the $<111>$ direction in the presence of a 1~ML step in VSS is evidenced through the orientation of the $(111)$ planes in the nanoparticles with respect to the same planes in the ZnTe nanowire. With no step present, the planes are parallel and the corresponding Au and ZnTe Bragg peaks are perfectly aligned in the FFT. This remains true in the presence of a 2~ML step (Fig.~\ref{fig3}a), as an effect of the small value of the coincidence mismatch. A clear misorientation by about 0.02~rad is detected in the presence of a 1~ML step (Fig.~\ref{fig3}b).

\begin{figure*}
 \centering
 \includegraphics[width=2\columnwidth]{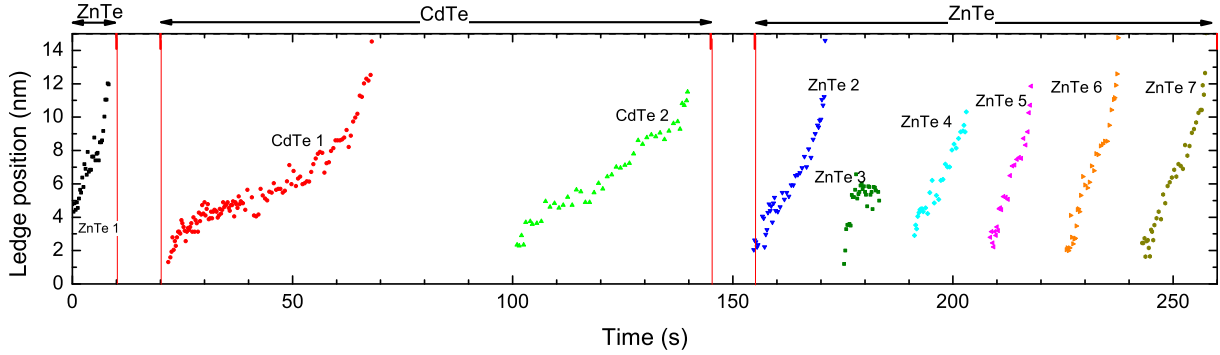}
  \caption{VSS growth: Position of the ledge as a function of time, under ZnTe or CdTe flux as indicated. All steps are 2~ML high. All started from the (same) nanowire side and moved regularly towards the opposite side, but ZnTe3 which followed a complex trajectory. NW 36-06, SiC substrate} \label{fig4}
\end{figure*}

We measured the dynamics of the steps on a series of nanowires where several successive steps could be observed precisely: start of the step (the first image showing the presence of the step), width of the island as a function of time, and end of the step (the last image where the step is visible). We define the cycle time for a given step, as the time between the end of the previous step and the end of the step under consideration. It comprises a “nucleation time” between the end of the previous step and the start of the step under consideration, and the “propagation time” between the start and the end of the step. The range of cycle times was 4 to 14~s for the 2~ML steps. The average of the cycle time on a given nanowire determines the growth rate (equal to 2~ML divided by the average cycle time) and it was found to be in direct relationship with the flux onto the gold nanoparticle (Supporting Information Sup1b).

The simplest case is that of a step starting on one side of the nanowire and propagating regularly, along a direction perpendicular to the electron beam, towards the opposite side of the nanowire. Then the island width is given by the position of the step ledge with respect to the nanowire side from which it started. Often however, the growth of a single nanowire combines steps with various directions of the step flow; Also, some steps exhibit complex trajectories with changing direction. For instance, in Fig.~\ref{fig2}e, a step starts from a side and propagates perpendicular to the e-beam for a while, then changes direction and reaches the front or backside. Such complex behaviors were established previously in the case of VLS-grown GaAs nanowires, using bird's eye views onto the gold-semiconductor interface of inclined nanowires \cite{Harmand2018}. The gold-ZnTe interface was imaged less precisely but the images (Supporting Information Sup1c) also suggest the possibility of more or less complex trajectories. Complex trajectories were frequently observed with a liquid droplet with a majority of 1~ML steps (Movie 1 in Supporting Information). A complex evolution of the island width as a function of time is generally the signature of a complex motion in the interface plane: An example will be given by step ZnTe3 in Fig.~\ref{fig4} below.

Figure~\ref{fig2}f shows for each studied sample the total cycle time as a function of nucleation time. In order to allow a comparison between different nanowires which receive different fluxes, the nucleation times and the cycle times measured on a given nanowire were divided by the average cycle time measured for the 2~ML steps on the same nanowire.

\begin{table}
  \caption{Average values and standard deviation for the nucleation, propagation and cycle times, for 1~ML and 2~ML steps. The times for each nanowire are normalized to the average 2~ML cycle time of the nanowire (hence the average 2~ML cycle time is unity by definition). The resolution is 1 frame (0.25~s) at each end; the cycle time ranges from 4 to 15~s with an average value of 9~s, the resolution is thus 0.06 times the average 2~ML cycle time.}
  \label{tbl:averages}
  \begin{tabular}{|c|c|c|c|}
    \hline
    Thickness  &          & Av. value & Stand dev.\\
    \hline
     & nucl   &0.25   & 0.12   \\
    1~ML   & prop   &0.30   & 0.11   \\
      & cycle   &0.55   & 0.09   \\
    \hline
     & nucl   &0.35   & 0.15   \\
    2~ML & prop   &0.65   & 0.16   \\
     & cycle   &1   & 0.10   \\
    \hline
  \end{tabular}
\end{table}

From Fig.~\ref{fig2}f we can conclude that:
\begin{itemize}
  \item The cycle times of 1~ML, 2~ML and 4~ML steps are in the ratio 1:2:4, corresponding to a single growth rate in ML/s.
  \item The nucleation is a random process and the nucleation time features huge fluctuations, from 10\% to 70\% of the total 2~ML cycle time: The standard deviation of the nucleation time is equal to 15\% of the total 2~ML cycle time (Table \ref{tbl:averages}). Although the nucleation time can be as low as 10\% of the cycle time, we never observed any well-characterized overlap of two independent 2~ML-steps.
  \item In spite of the strong fluctuations in the nucleation time, the cycle time of 2~ML-steps within each nanowire is close to the average (i.e., close to unity in Fig.~\ref{fig2}); The standard deviation is only 10\% (Table \ref{tbl:averages}), much lower than expected for independent nucleation and propagation times. There is no sizable dependence on the nucleation time, the propagation time adjusts and smoothes the fluctuations of the nucleation time.
  \item The very few 1~ML steps that were observed in VSS feature a very short propagation time. Considering both VSS and VLS, a regulation is observed also for the 1~ML steps but it is less efficient than for the 2~ML steps.
\end{itemize}

A deeper understanding of the step dynamics is achieved by plotting the position of the moving step as a function of time (Fig.~\ref{fig4}). We focus on 2~ML steps which move more slowly than the 1~ML steps. The sequence comprises a short ZnTe section, a 10~s growth interruption, a CdTe section, another 10~s growth interruption, and a final ZnTe section (see also Movie 5 in Supporting Information). We have chosen this sequence because all steps started from the same side of the nanowire, and all but ZnTe3 moved regularly towards the opposite side. All are 2~ML steps. In the following we discard step ZnTe3.

\begin{figure*}
 \centering
 \includegraphics[width=2\columnwidth]{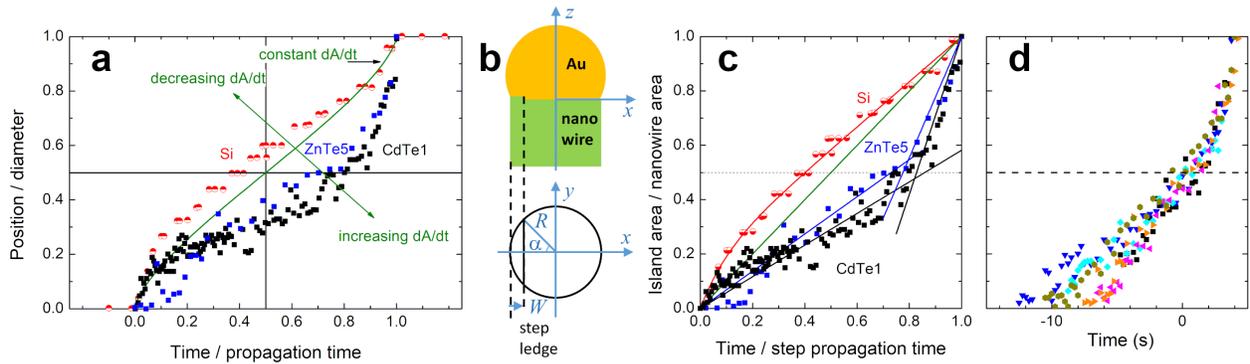}
  \caption{ (a) position of the ledge as a function of time (both scales in reduced units, diameter and propagation time, respectively) in Si (red half-filled circles) \cite{Ross2010,Cui2015}, and CdTe1 (black squares) and  ZnTe6 (blue squares) of Fig.~\ref{fig4}; The position of a step moving perpendicular to the electron beam at constant $dA/dt$ across a circular nanowire is shown by the green solid curve. (b)  Simple scheme of a step across a circular nanowire. (c) Corresponding evolution of the island area $A(t)$. The red curve is a fit with an initial excess decreasing exponentially; The black (blue) lines are a linear fit of the CdTe (ZnTe) data below and above half completion, respectively. (d) Comparison of the six ZnTe steps of Fig.~\ref{fig4}; For each step, the origin of the horizontal scale is the time for half filling, estimated from the average of 4 to 6 data points around half filling.} \label{fig5}
\end{figure*}

Figure~\ref{fig5}a plots the evolution of steps CdTe1 and ZnTe6 from Fig.~\ref{fig4}, in reduced coordinates. We observe an S-shape of the step propagation, as in other systems previously described: Figure~\ref{fig5}a gives an example for Si nanowires \cite{Ross2010}. In this case an asymmetry of the S-shape was observed: the step velocity is larger when the step starts than when it reaches the opposite side, so that more than half of a monolayer has formed at mid propagation time. In our case, the asymmetry is opposite, with a slower initial motion and a faster final motion, and less than half-completion at mid propagation time. The example of Fig.~\ref{fig5}a is confirmed for all steps in Fig.~\ref{fig4} and it is a general feature in the present study.

In the simplest picture, a linear step moving in the direction perpendicular to the electron beam across a circular nanowire of radius $R$ (Fig 1 of Ref.~\citenum{Cui2015} and the schematic in Fig.~\ref{fig5}b), the area of the semiconductor island is $A=R^2 [\alpha-\frac{1}{2}\sin 2 \alpha]$  and its width (\emph{i.e.}, the position of the step ledge) is $W=R (1-\cos \alpha$), hence $\frac{A}{\pi R^2} = \frac{1}{\pi} [\arccos(1-\frac{W}{R})-(1-\frac{W}{R})\sqrt{1-(1-\frac{W}{R})^2})]$. If $dA/dt$ is constant, $A=\frac{dA}{dt} t$ and we expect a symmetric displacement $W(t)$, solid green curve in Fig.~\ref{fig5}a and solid green straight line in Fig.~\ref{fig5}c. It actually coincides with the result of the detailed diffusion model shown in Fig. 3 of Cui \emph{et al}.\cite{Cui2015}.

The asymmetric dynamics observed on Si NWs \cite{Ross2010}, with a faster start, was attributed to an initial excess of the Si content. This is well exemplified if we plot the area of the island $A(t)$ instead of the step position $W(t)$ (Fig.~\ref{fig5}c). In Ref.~\citenum{Ross2010}, the initial excess was assumed to vanish almost immediately, leaving the concentration at its steady-state value. We have calculated the solid red line in Fig.~\ref{fig5}c assuming that the initial excess decreases with a characteristic time equal to 1/10 the cycle time: This assumption, albeit fully phenomenological, reproduces surprisingly well the experimental data (shown by symbols). However, it cannot describe our results on ZnTe or CdTe. In this case, we observe a similar effect at very short time, but finally the overall asymmetry is opposite: The step speeds up at half-filling, and the $A(t)$ plot definitely features a slow slope below half-filling and a larger one above.
Finally, Fig.~\ref{fig5}d compares the dynamics $A(t)$ of the six regular ZnTe steps of the nanowire of Fig.~\ref{fig4}. The origin of the horizontal scale is the time for half-filling ($A(t)=\frac{1}{2}\pi R^2$, $W(t)=R$). A striking feature is the strong dispersion of the left-hand side, while all steps display more or less the same dynamics on the right hand side: That means that the fluctuations of the nucleation time are compensated already at half-filling.

\subsection{Formation of a CdTe-ZnTe interface}

The congruent nature of the evaporation of CdTe and ZnTe gave us the opportunity to study the formation of an interface between two compound semiconductors, using only two evaporation cells.

Coming back to Fig.~\ref{fig4}, we have already noted that the step motion under CdTe or ZnTe flux appears similar when scaled to the propagation time. However, the most striking observation is the rapid change of the timescale of dynamics (step propagation, and hence growth rate). From the position of the nanowire with respect to the CdTe and ZnTe cells, and from the cell temperatures, we estimate the CdTe flux to the nanoparticle to be about four times smaller than the ZnTe flux (see Supporting Information S1b). This is the order of magnitude observed for the velocity of the two steps under CdTe flux, with respect to the step velocity under ZnTe flux. Note that the change of velocity is observed on the very first steps after the reestablishment of a flux following the growth interruption.

We may note the remarkably short nucleation time when the ZnTe cell is opened after the interruption of the CdTe flux: Step ZnTe2 starts immediately and propagates slowly with respect to the other ZnTe steps, and indeed it is the extreme case in Fig.~\ref{fig5}d. Nevertheless, the dynamics during the second half of the propagation is similar to that of the other ZnTe steps.

Unfortunately, our attempts to extract the interface profile from the local interplane distance were not successful. To be significant, a geometrical phase analysis (GPA) would require to be performed on images with a larger scale recorded on a longer CdTe sequence.

\subsection{Discussion}

We observed a change of shape of the solid nanoparticle during the growth of the nanowire, from facetted to almost spherical. This is possibly induced by a change of surface energy due to the incorporation of Zn, homogeneous or not. During the growth of ZnO nanowires at 1173K, the presence of a Zn-rich liquid layer at the surface was invoked \cite{Simon2013}. Indeed the surface energy in the same configuration, the (111) surface of fcc Au and the (0001) surface of bcc Zn, is 0.611 eV/atom for Au and 0.385 eV/atom for Zn \cite{Vitos1998}. This large difference could favor a Zn-rich surface, possibly up to the eutectic concentration. However, in the present case we observe that the lattice planes reach the nanoparticle surface and this rules out the formation of a thick liquid layer at the surface. Another parameter would be the surface energy of the alloy, but the available information on the Au-Zn alloy is too scarce to allow any meaningful discussion. In the present study, the change of shape was observed in coincidence with the presence of a step:  indeed, the nucleation and the propagation of the steps are expected to impact the Zn concentration and its distribution in the gold nanoparticle, and thus play a role in the change of shape, but this has to be confirmed. Finally, we have described earlier \cite{Rueda2016} the role of the nanoparticle shape in the growth rate of ZnTe nanowires under standard MBE growth conditions, in relation with a change in the nanoparticle-nanowire contact area: If this is induced by the change of shape observed in the present study, the change in contact area does not occur suddenly and it is not detected on the short time scale of the present TEM movies.

A key ingredient of the gold-seeded VSS growth of ZnTe appears to be the mismatch between the Au and ZnTe lattices, and the near-coincidence at 2:3 ratio. The relative orientation of the Au and ZnTe lattices realizes the near-coincidence within the (111) plane. We now show that the lattice mismatch is also responsible for the tilt of the (111) planes in the presence of a 1~ML step (and not for a 2~ML step where the near coincidence is achieved also along the $\langle111\rangle$ direction), and that it leads to a significant elastic energy which favours the 2~ML steps at the expense of the 1~ML steps.

The strain state of the system with a partial ZnTe ML was computed in the framework of linear elasticity by using the finite elements software COMSOL \cite{COMSOL}. We consider a step of height 1~ML (=0.352~nm) at the interface between an axisymmetric ZnTe NW of radius $R$ and a spherical cap of solid Au of height $H$. The axes of the two cubic crystals are parallel, with [111] along the cylinder axis $z$. The straight step lies along the [11$\bar{2}$] $y$ axis and separates the ZnTe part of the ML, which extends over width $W$ along the [1$\bar{1}$0] $x$ axis, see Fig.~\ref{fig5}b. The stiffness tensors of the two materials were calculated in these axes \cite{FerandStrainCS2014} from the elastic constants of ZnTe\cite{LeeZnTeStiffness} and Au\cite{ChangAuStiffness}. We assume that the two crystals are coherently assembled over their entire interface, with exact in-plane lattice coincidence over the horizontal terraces on either side of the step (the small coincidence mismatch is neglected), whereas the misfit in the vertical step interface (where coincidence is impossible), is that between bulk ZnTe and Au, $(a_0^{ZnTe} - a_0^{Au})/ a_0^{ZnTe} = -0.332$. Writing $u_{x,y,z}$ the displacements along the three axes, Fig.~\ref{fig3}c displays the shear strain $\frac{1}{2}(\frac{\partial u_z}{\partial x}+\frac{\partial u_x}{\partial z})$, Fig.~\ref{fig3}d the rotation $\frac{1}{2}(\frac{\partial u_z}{\partial x}-\frac{\partial u_x}{\partial z})$, and Fig.~\ref{fig3}e the tilt of the (111) planes in the $xz$ axial plane, $\frac{\partial u_z}{\partial x}$, for a half ZnTe ML ($W = R$), in the case $R = 3.6 \textrm{ nm}$, $H = 6.4 \textrm{ nm}$. These maps, all in the same color scale, show that the strain is strong and inhomogeneous around the step but extends only over a few atomic spacings. Over the main part of the nanoparticle and nanowire, $\frac{\partial u_z}{\partial x}$ and $\frac{\partial u_x}{\partial z}$ almost cancel and the shear strain almost vanishes. We thus have essentially a rigid-body rotation of the nanoparticle (Fig.~\ref{fig3}d) with uniform tilt of the (111) planes (Fig.~\ref{fig3}e). Finally, Fig.~\ref{fig3}f shows the histogram of $\frac{\partial u_z}{\partial x}$ over the whole Au nanoparticle, with a maximum at 0.021 rad, in good agreement with the experimental value.

Can we understand this value of the tilt? This rigid body rotation which persists far away from the step is similar to the effect of dislocations at a grain boundary \cite{HirthLothe}. Then the angle between the two grains is $b/2R_0$, where $b$ is the length of the Burgers vector (or its projection on the normal to the interface), and $2R_0$ the distance between dislocations. The elastic energy per unit length of an edge dislocation in bulk, as given in textbooks \cite{Kittel}, is $E=\frac{Gb^2}{4\pi(1-\nu)}\ln(\frac{R_0}{r})$, with $G$ and $\nu$ the shear modulus and Poisson ratio describing isotropic linear elasticity. The two characteristic lengths $R_0$ and $r \simeq b$ play a moderate role since they appear through the logarithm. The proportionality to the shear modulus and to the squared Burgers vector are a direct consequence of applying linear elasticity. The same expression was proposed in Ref.~\citenum{Glas2006} for a dislocation at the interface of an axial nanowire heterostructure, with $r=b$ and $R_0$ being the average distance to the nearest free surface. In the present case, this leads us to define an effective Burgers vector of length $b=(d_{ZnTe}-d_{Au})=0.117$~nm, acting over a distance equal to the nanowire diameter $2R$. In Fig.~\ref{fig3}b, the diameter is 7.2~nm, hence the expected order of magnitude of the tilt angle is $b/2R=0.016$~rad, in good agreement with the measured angle of $0.02$~rad and with the result of the COMSOL calculation (which also predicts a decrease of the tilt angle by a factor of about 2 on a twice broader nanowire).

At a 2~ML step, coincidence in the vertical plane becomes possible (with again the small coincidence mismatch $f=-0.3\%$ given above). In practice, the rigid-body rotation disappears (Fig.~\ref{fig3}a).

A closely related calculation was performed in Ref.~\citenum{Koryakin2019a} and \citenum{Koryakin2019b} in order to address the VSS growth of GaAs or InAs seeded by (Au,Ga) or (Au,In) solid solutions or intermetallic compounds. The authors calculate the strain induced by an island at the semiconductor-seed interface by a variational method, and show that the energy contains three contributions, an interphase energy (estimated to be a few 10~meV per InAs pair), an elastic term proportional to the step length, and an elastic term proportional to the island area. They finally compute, for different compositions of the seed, the energy of the critical nucleus. The elastic energy varies from 8 to 441 meV per InAs pair when the lattice mismatch varies from $f=0.02$ to 0.17, and one can easily check that the dependence on $f$ is quadratic, as expected from linear elasticity.

The present system differs from the system of Ref.~\citenum{Koryakin2019a} and \citenum{Koryakin2019b} since it involves a good coincidence in the interface plane, instead of a mismatch along the three directions. The elastic energy calculated with COMSOL for the structure of Fig.~\ref{fig3}, with the 1~ML step centered, remains high, 0.8~eV nm$^{-1}$. The elastic energy at the 2~ML step, proportional to the square of the mismatch $f=-0.3\%$, is negligibly small. We may conclude that the energy of the 2~ML critical nucleus is mainly given by the interphase energy, while that of the 1~ML one involves a strong elastic energy. This implies that the nucleation time of the 1~ML step in VSS is expected to be long (in other words, the probability of a 1~ML step is small), while the nucleation time of the 2~ML step involves a critical nucleus energy of the same order as in VLS. Indeed, almost no 1~ML steps are observed in the VSS growth of ZnTe, while they are present in VLS.

The step energy at the Au-semiconductor interface is still a poorly known quantity, both theoretically and experimentally. In the context of VLS, it is quite often approximated by the surface energy obtained for a large area, and calculated for an area determined by the ledge length and the step height. Hence for identical area of the critical nucleus, the 1~ML step is strongly favoured. In VSS, as a first approximation, we may considered that the energies are enhanced by the elastic contribution: Indeed, the VSS growth of GaAs displays both 1ML and 2ML steps \cite{Maliakkal2021arX}, suggesting that the larger surface energy of the 2ML step is compensated by the stronger elastic energy of the 1ML step.

The VLS growth of ZnTe involves both 1ML and 2ML steps. This is not expected for a liquid seed, and implies that either the 1~ML and 2~ML critical nuclei have different shapes, or that the step energy is not proportional to the step height. We note that some ordering exists in the liquid gold seed, close to the interface with the semiconductor (see Movie 1), as often observed in other systems. This ordering may play a role in the energy of the critical nuclei and its formation.

The same 3:2 ratio exists for Au:CdSe, InAs or GaSb, although the picture can be altered by the formation of an alloy with a large In or Ga content hence a different lattice parameter. One may note also that Au:Si offers a ratio close to 4:3. Similarly, ZnO nanowires were grown at low temperature (350$^\circ$C) on crystalline, lattice matched $\gamma$-AuZn nanoparticles \cite{Campos2008}. In the present study the Zn content is low and the lattice mismatch is minimized for (almost) pure Au and ZnTe. We note that 2~ML steps are also obtained on the few CdTe steps we have observed in spite of a larger coincidence mismatch ($\simeq 6\%$) so that the elastic energy at the 2~ML step is not totally negligible with respect to the interphase energy: a more systematic study is needed in this case.

The most interesting consequence is the stabilization of the cycle length, corresponding to an adjustment of the growth rate to the available flux. The random character of the nucleation is almost totally washed out. It was noted early \cite{Glas2010} that the depletion which follows the nucleation and the rapid expansion of the island at the interface leads to an antibunching of the nucleation process. The characteristics of this depletion was analyzed further for a low incident flux \cite{Dubrovskii2017,Glas2020} thanks to the quantitative approach which can be developed for the VLS growth of III-V nanowires. In this case a basic process is thought to consist of two steps: (1) the increase of the concentration of group V atoms in the droplet until random nucleation and rapid expansion of the island fed by the excess of concentration, and (2) a final expansion to  the full ML controlled by the incoming flux. During the second step, a steady-state concentration is reached as a result of the balance between incoming flux and incorporation at the propagating step. Hence, the concentration is the same at completion of all MLs. The excess concentration in the nanoparticle accumulated before the random nucleation event is incorporated entirely into the growing nanowire during the propagation of the step, so that in the absence of desorption (an assumption which was checked to hold for Zn and to break down for Cd only at temperatures above 300$^\circ$C \cite{Orru2018}), the cycle time is fixed independently of the random character of nucleation.

The stabilization demonstrated in the global analysis (Fig.~\ref{fig2}f) and Table~\ref{tbl:averages} points to a similar mechanism and explains why the growth models based on the available flux and omitting the nucleation process \cite{Rueda2016} can be used. Note that the 1~ML steps, which propagate faster, feature a less stable cycle length, which still remains close to half the value of the 2~ML cycle time. Indeed, it is worth noting at this point that when plotting the growth rate as a function of the incident flux (see Supporting Information Sup1b), we did not observe any significant difference between VSS and VLS. The only deviation is observed after the recrystallization of the liquid droplet into two separate Au and Si nanocristals.

The detailed analysis (Figs. ~\ref{fig4} and \ref{fig5}) suggests some modifications with respect to this two-step process. From fig.~\ref{fig5} we can conclude that:
\begin{itemize}
  \item The early expansion of the island – which consumes the excess concentration accumulated before nucleation – may appear as a fast initial propagation of the step, before the stabilization.
  \item An additional final step is observed between half coverage and full coverage, with a faster propagation of the step. A possible interpretation is a slow build-up of the step followed by a fast erasing, to be analyzed with respect to the cost in energy: The cost is larger when the step length increases during the first half of step propagation than when it decreases during the second half. A detailed description of the role of step energy is given in Ref. \citenum{Glas2020}. A result of the present study is that the $A(t)$ behavior is more complex than those predicted by current models of VSS \cite{Golovin2008,Cui2015}.
\end{itemize}

The standard deviation for the cycle time of 2~ML steps is well below the combined standard deviations of nucleation and propagation assumed to be independent, see Table 1, but it remains above the standard deviation expected from the resolution of 0.25~s (at both the step start and the step end). Here again an influence of the detailed morphology of the steps, and of the occurrence of complex trajectories, should be studied in detail. For instance, even for a simple scheme, in the zinc-blende structure, the arrangement of atoms at a step propagating from left to right is different from the arrangement at a step propagating from right to left.

The change of step velocity at the ZnTe to CdTe interface can be due either to a different flux intensity when switching from ZnTe to CdTe, or to re-evaporation of one constituent. Indeed, we know from a previous analysis of MBE under standard conditions \cite{Orru2018} that Cd is quite volatile and evaporates from the Au nanoparticle at temperatures above $300^\circ$C. This is in principle higher than the substrate temperature in the present study. The second mechanism discussed in Ref.~\citenum{Orru2018}, sublimation from the nanowire sidewalls, does not play any role here since the contribution from the sidewalls is already quenched by the electron beam (see Supporting Information Sup1b). Hence we ascribe the change of dynamics to the low CdTe flux already mentioned. In Figs.~\ref{fig4} and \ref{fig5}, step ZnTe2, the first one after the interruption of the CdTe flux and the reestablishment of the ZnTe flux 10~s after, exhibits a slow initial propagation time: this may be either an effect of the flux interruption, or an effect of the low CdTe flux. However, the fast dynamics is fully recovered for the second half of the propagation (Fig.~\ref{fig5}d). Such a fast recovery must be taken into account when modelling the (Cd,Zn) profile across the interface. Previous interfaces were estimated to be about 2~nm thick \cite{Orru2018}, much thicker than the height of a 2~ML step (0.7~nm), but they were grown with a larger growth rate, hence a higher content in the gold nanoparticle, and without growth interruption. Hence, we can expect a smaller reservoir effect under the present conditions.

\section{Conclusions}

To sum up, the gold-seeded growth of ZnTe nanowires under conditions close to that involved when growing on a ZnTe buffer layer is confirmed to take place in the Vapor-Solid-Solid (VSS) mode. The whole gold nanoparticle exhibits crystal planes, although its shape can be either facetted or almost spherical, and even change during the growth. The growth is ensured essentially through the nucleation and propagation of 2-ML steps, in agreement with the lattice coincidence which exists between the zinc-blende lattice of ZnTe and the fcc lattice of Au: a 2ML step involves a very small mismatch, in contrast with the large mismatch strain associated with the 1ML step. By contrast, we observed that the VLS growth involves a mixing of 1~ML and 2~ML steps, with no decisive change in the growth rate since the cycle time for a 1~ML step is about half that for a 2~ML step. However the recrystallization of the liquid droplet formed on a silicon substrate may result in a Si-Au phase separation, with an impact on the growth rate.

The growth regime we have explored in the present study is characterized by a partial self-regulation of the cycle (nucleation + propagation) time, in agreement with models developed for the case of a low incident flux: The strong dispersion of the nucleation time is partially compensated by the propagation time. However, the details of the step propagation point to a different incorporation rate at the step edge during its propagation even if the step apparently follows a simple trajectory.

\begin{small}

\section{Methods}
Growth experiments are performed by molecular beam epitaxy in a modified FEI environmental TEM operated at 300 keV and equipped with an image aberration corrector. The background pressure in the object chamber is $8\times10^{-6}$ Pa. Two external ports are used to fit thermal effusion cells loaded with ZnTe and CdTe, respectively. The vapor fluxes reach the sample holder through collimators having an internal diameter of 1 mm. Over the range of cell temperature we use, the sublimation of these compounds is congruent. Therefore, the VI/II flux ratio is always equal to one during growth. Nanowire formation is catalyzed by Au particles dispersed on a micro-heater substrate made of silicon or silicon carbide. During the growth, the substrate temperature is estimated to be in the low 300°C range. The growth rate is typically a few (2 to 8) nm/min, and it is proportional to the flux directly impinging the nanoparticle calculated from the NP shape and orientation with respect to the flux (see Supporting Information Sup1b). This growth rate is smaller than in a standard MBE setup (few tens of nm/min), where a contribution from the flux to the sidewalls over a diffusion length approx. 100 nm (ratio sidewall / nanoparticle contributions about 5) is evidenced \cite{Rueda2016}. This sidewall contribution appears to be quenched in the present study. Indeed the electron beam has a visible effect on the sidewalls of the nanowire, and causes a desorption of the adatoms and even a sublimation of the nanowire after several minutes.

\begin{acknowledgement}
We acknowledge funding by the French National Research
Agency  (project ESPADON ANR-15-CE24-0029, and NanoMAX equipment 10-EQPX-0050 TEMPOS (Transmission Electron
Microscopy at Palaiseau-Orsay-Saclay)), and CIMEX at \'{E}cole polytechnique (Palaiseau) for
hosting this equipment.

\end{acknowledgement}

\begin{suppinfo}

The following files are available free of charge.
\begin{itemize}
  \item Sup1: (a) \emph{ex-situ} analysis of the orientation of the gold nanoparticles. (b) The 2ML cycle time (average) and the flux onto the nanoparticle. (c) Evolution of an island on a tilted interface, nanowire 72-26. (d) List of videos.
  \item Movie 1: nanowire 13-25, VLS, ZnTe on Si substrate
  \item Movie 2: nanowire 35-26, VSS, ZnTe on Si substrate
  \item Movie 3: nanowire 13-25, ZnTe on Si substrate; solid core moving in a liquid droplet.
  \item Movie 4: nanowire 31-06, VSS, ZnTe on SiC substrate; with steps moving along the e-beam, and shape change.
  \item Movie 5: nanowire 36-06, VSS on SiC substrate; ZnTe and CdTe with steps moving perpendicular to the e-beam.

\end{itemize}

\end{suppinfo}


\end{small}

\end{document}